\documentclass[useAMS,usenatbib]{mn2e}

\usepackage{aas_macros}
\usepackage{dcolumn}
\usepackage{amsmath}
\usepackage{amssymb}
\usepackage{longtable}
\usepackage{rotating}
\usepackage{graphicx}
\usepackage{subfigure}

\newcommand{\Msun}{M$_\odot$}

\newcommand\mean[1]{\overline{#1}}
\newcommand{\degree}{\ensuremath{^\circ}}

\newcommand{\sigz}{$\sigma_0$}

\begin{document}
\title[The central mass and M/L profile of M15]{The central mass and mass-to-light profile of the Galactic globular cluster M15}
\author[M. den Brok et al.]{Mark~den~Brok,$^{1,2}$\thanks{denbrok@physics.utah.edu} Glenn~van~de~Ven,$^3$, Remco~van~den~Bosch,$^3$,  Laura~Watkins$^3$\\
$^1$Kapteyn Astronomical Institute, University of Groningen, P.O. Box 800, 9700AV Groningen, The Netherlands.\\
$^2$Department of Physics and Astronomy, University of Utah, Salt Lake City, Utah 84112, USA\\
$^{3}$Max Planck Institute for Astronomy, K\"onigstuhl 17, 69117 Heidelberg, Germany\\
}
\maketitle
\begin{abstract}
We analyze line-of-sight velocity and proper motion data of stars in the Galactic globular cluster M15 using a new method to fit dynamical models to discrete kinematic data. Our fitting method maximizes the likelihood for individual stars and, as such, does not suffer the same loss of spatial and velocity information incurred when spatially binning data or measuring velocity moments.
In this paper, we show that the radial variation in M15 of the mass-to-light ratio is consistent with previous estimates and theoretical predictions, which verifies our method. Our best-fitting axisymmetric Jeans models do include a central dark mass of $\sim2 \pm 1\cdot 10^3M_\odot$, which can be explained by a high concentration of stellar remnants at the cluster center. This paper shows that, from a technical point of view and with current computing power, spatial binning of data is no longer necessary. This not only leads to more accurate fits, but also avoids biased mass estimates due to the loss of resolution. Furthermore, we find that the mass concentration in M15 is significantly higher than previously measured, and is in close agreement with theoretical predictions for core-collapsed globular clusters without a central intermediate-mass black hole. 
\end{abstract}

\section{Introduction}
\label{sec:intro}
M15 is well known as the prototypical core-collapsed globular cluster \citep{DjoKin86}. Unlike 'normal' globular clusters with constant central luminosity densities ($\sim80$\% of the Galactic globular clusters) the light profile of a  core-collapse cluster rises all the way to the center. Core-collapse is supposed to be a result of a gravo-thermal catastrophe, caused by the negative heat capacity of gravitational systems \citep{Ant62,LynWoo68}. Mass segregation in these system is responsible for a high fraction of neutron stars and white dwarfs near the center of the cluster. Indeed a high number of pulsars is observed in M15 \citep{Phi93}, though almost all of them outside the core.

Globular clusters are interesting places to search for intermediate mass black holes (IMBHs) \citep{Wyl70}. Na\"ively, one might expect a core-collapsed globular cluster to be a likely host for an IMBH as their formation has been linked to the runaway growth of stars \citep[see e.g.][]{QuiSha90}. Core-collapse, however, can be halted by the addition of energy to the core from 'binary burning' \citep{GooHut89}, so that the densities for runaway growth are not reached. High-mass stars can undergo core-collapse independently of the rest of the cluster, because energy equipartition does not necessarily hold. The N-body simulations of \citet{PorMcM02} produce IMBHs from core-collapsing high mass stars during the early stages of the cluster's life time. This is supported by the models of \citet{GurFreRas04}, who, however, state that the initial concentration of M15 should have been much higher for the run-away growth of a black hole from stellar mergers. Moreover, the black hole itself can also function as a heating mechanism to halt further collapse, because, as captured stars are likely the ones carrying little energy, their removal leads to an increase of the average kinematic temperature \citep[e.g.][]{HutMcMGoo92}. Finally, \citet{BauMakHut05} have shown using N-body simulations that, since an IMBH would quickly puff-up the core, centrally concentrated clusters such as M15 are in fact the least likely GCs to host an IMBH.

From an observational point of view, the presence of IMBHs in globular clusters has not been settled. Besides for M15, there have been kinematic detections of IMBHs in, among others, G1 \citep{GebRicHo05}, $\omega$ Centauri \citep[][but see \citealt{vanAnd10}]{NoyGebKis10} and NGC6388 \citep[][but see \citealt{LanMucOri13}]{LutKisNoy11}. However, non-kinematic signatures of IMBHs in clusters are often absent \citep[e.g.][]{StrChoMac12,MilWroSiv12}

Located at a distance of 10.4 kpc \citep{DurHar93}, M15 has been extensively studied in the past. Several IMBH measurements for this cluster are reported in the literature \citep{GebPryWil97,GerVanGeb02}. The most recent estimate comes from \citet[][hereafter vdB06]{vandeZGeb06}, who modeled the cluster with axisymmetric Schwarzschild models. The preferred dark central mass found by these authors was 500\Msun, although the absence of an IMBH could not be ruled out.

Here we re-analyze the same data used by vdB06 (Sec. \ref{sec:data}), though with a different approach. We use axisymmetric Jeans models to constrain the mass distribution, but refrain from binning the data, which gives us higher resolution, in particular in the center of the cluster. Although discrete fits to the internal kinematics of GCs have been performed before \citep[e.g.][]{IbaNipSol13}, our method with axisymmetric models allows for a non-spherical mass distribution (Sec. \ref{sec:modeling}). The purpose of this paper is two-fold: first, we show that the statistical method which we developed gives physically meaningful results when applied to real data (Sec. \ref{sec:bestfit}). Second, we show that the mass concentration in the center of M15 is significantly higher than that obtained with the dynamical modelling of binned data (Sec. \ref{sec:bh}). We discuss our results and present our conclusions in Sec. \ref{sec:conc}.

\section{Data}
\label{sec:data}
\subsection{Line-of-sight velocities}
We use the measured line-of-sight (LOS) velocities of \citet[][hereafter G00]{GebPryOCo00}, based on adaptive optics assisted Fabry-Perot measurements with the Canada-France-Hawaii Telescope. Following vdB06, we restrict ourselves to the high-quality subset of 1546 stars from the initial sample of 1773 stars, for which the errors on the velocity are smaller than 7 km/s. We also include the 64 stars from the observations with STIS on board the Hubble Space Telescope (HST) by \citet[][hereafter vdM02]{vanGerGuh02}. As they are close to the center of the cluster, they provide an important constraint on the mass of a possible IMBH. Most of the stars ($>80$\%) in the LOS velocity sample are located within the central 1.5', but the data extend as far as 15'.
\subsection{Proper motions}
\begin{figure*}
  \begin{center}
    \includegraphics[angle=0, width=.48\textwidth]{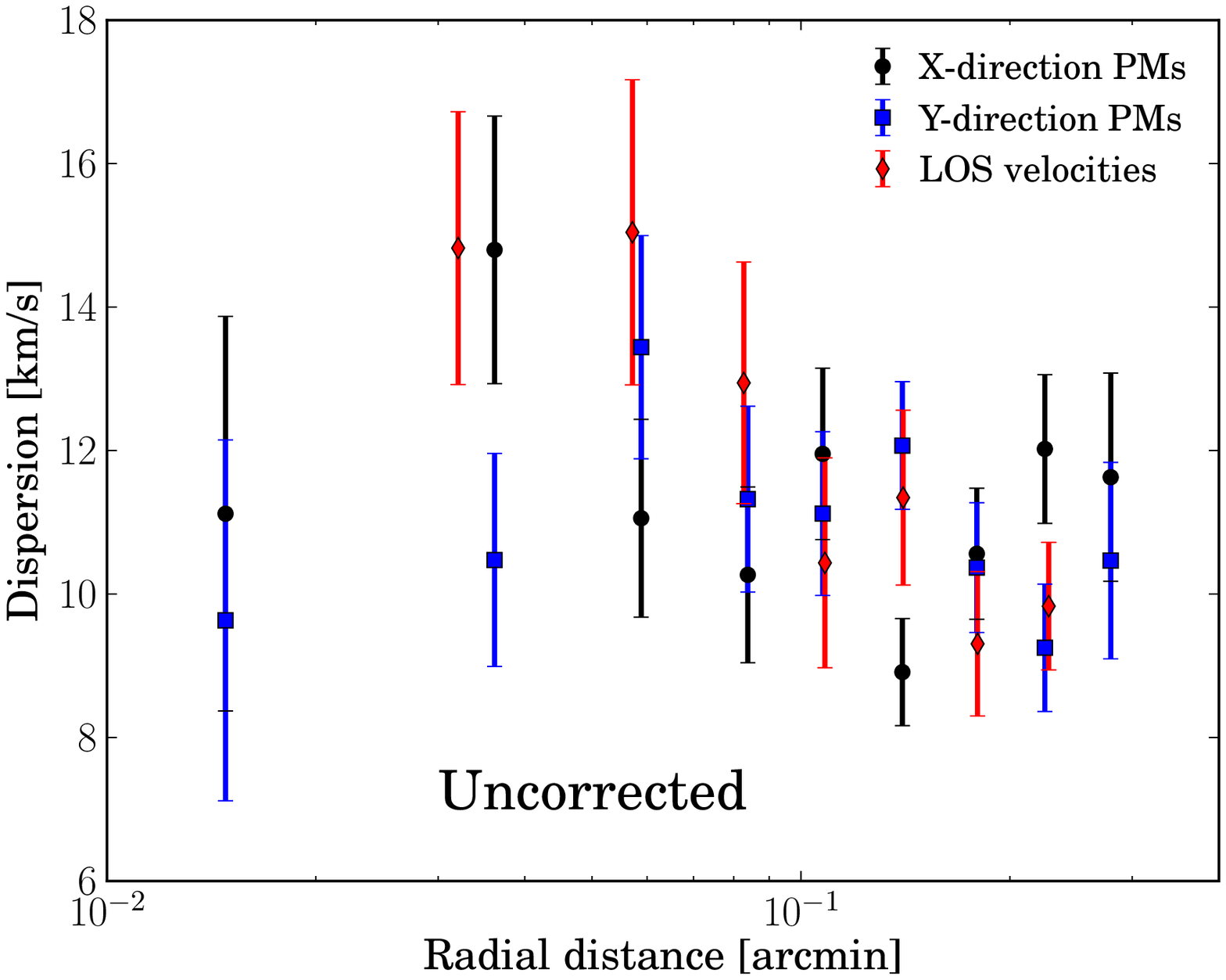}
    \includegraphics[angle=0, width=.48\textwidth]{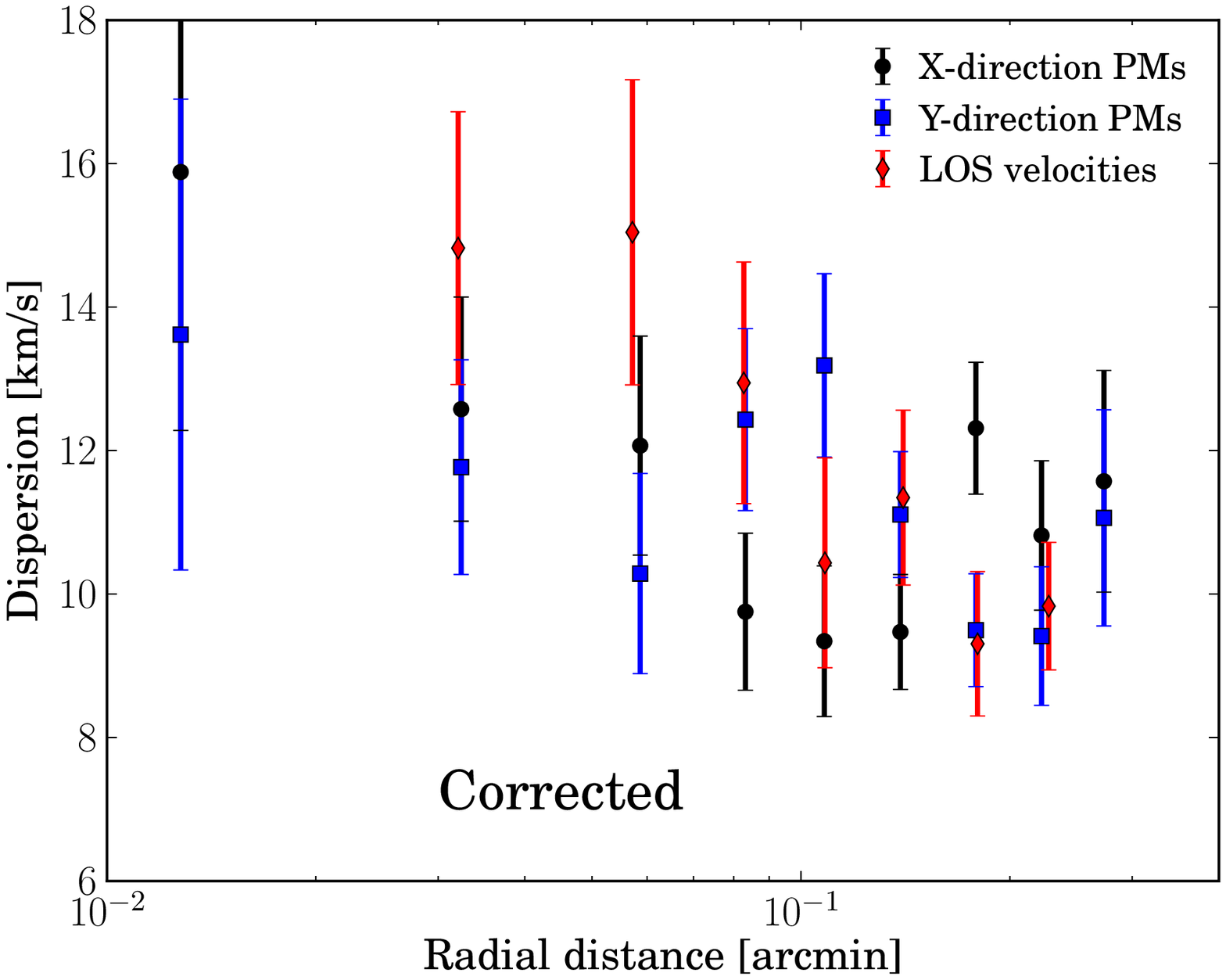}
  \end{center}  
  \caption{Radial dispersion profiles for M15. Shown are the LOS velocity (red diamonds) and proper motion dispersons in the X and Y direction (black circles and blue squares). In the right panel, the proper motion data have been corrected for the offset in stellar position with respect to the line-of-sight data.} 
  \label{fig:dispersions}
\end{figure*}
\citet[][hereafter M03]{McNHarAnd03} measured proper motions (PMs) of 1764 stars from HST/WFPC2 images. Here we use the same subset as used by vdB06, i.e., the 704 stars with B magnitude brighter than 16.5. During the course of this 
work we discovered an offset between the positions of the stars as given in M03 and the stars in the sample of vdM02. We correct the PM data, resulting in slightly different dispersion profiles, as shown in Fig. \ref{fig:dispersions}. To convert observed angular motions into physical velocities, we adopt a distance to the cluster of 10.4 kpc \citep{DurHar93}. The PM data cover the central 0.3', and thus can provide an additional constraint on the central mass distribution. 

\subsection{Spatially binned data}
In order to compare our likelihood method with results based on spatially binned data, we generate a binned data set for both PM components and the LOS velocity component.
For each component, we reflect the x-y positions of our data so that all stars lie in one quadrant. After that, we construct polar bins in a similar way as vdB06, by dividing this quadrant into radial shells, which are subsequently divided into 1, 2 or 3 angular bins. Although the radial shells are not exactly the same as in vdB06, the division of radial shells into angular bins has the same radial dependence as in vdB06. In each bin, we determine the intrinsic dispersion \sigz, taking into account the measurement errors, by maximizing the likelihood of the stars in the bin: 
\begin{eqnarray}
\mathcal{L} = \prod_i\frac{1}{\sqrt{2\pi(\sigma_0^2 + \delta v_i^2)}}\exp\left( -\frac{v_i^2}{2(\sigma_0^2 + \delta v_i^2)}\right),
\end{eqnarray}
where $v_i$ and $\delta v_i$ are the observed velocity and velocity error of a star in a PM or LOS direction. The small amount of rotation (vdB06) is neglected. The error on \sigz\ is then found by inverting the Fisher information on \sigz.

\section{Dynamical modelling}
\label{sec:modeling}
We fit the data with axisymmetric anisotropic Jeans models, using a modification of the code from \citet{Cap08}. To solve the Jeans equations, the code uses a Multi-Gaussian Expansion (MGE) of both the projected mass density and the light distribution and requires the evaluation of only one numerical quadrature per predicted second moment. In addition to the second moment of the velocity along the line-of-sight, the code also predicts the second moments in both PM directions, based on the derivations of \citet{Watvanden13}.

\subsection{Discrete fitting}
Since the observations consist of velocities of individual stars, we calculate, for a given mass distribution, the velocity distributions at the position of each individual star. The difference with respect to earlier modelling is that we do not bin our data, but calculate the total likelihood by taking the product of the likelihood of the velocity of each individual star. The Jeans models used to predict the velocities are, however, less sophisticated than the dynamical models of vdB06, because they contain strong assumptions on the velocity anisotropy.

 If $v_{i} \pm \delta v_{i}$ is the observed velocity plus error of a star, we can calculate the likelihood for this star as:
\begin{eqnarray}
\mathcal{L}_i = \frac{1}{\sqrt{2\pi \left(\mean{v_{i,mod}^2} + \delta v_{i}^2 \right)}} \exp{\left(-\frac{1}{2}\frac{v_{i}^2}{\mean{v_{i,mod}^2} + \delta v_{i}^2} \right)},
\end{eqnarray}
where $\mean{v_{i,mod}^2}$ is the prediction of the second velocity moment from the Jeans model. As before, we neglect the small amount of rotational motion (vdB06), so that the first velocity moments vanish.
Our modelling has advantages over previous modelling, which has nearly always been based on spatially binning data to calculate the second moment of the velocity \citep[but see][]{ChaKlevan08}. Avoiding binning gives us higher spatial resolution close to the the center of the cluster. A more elaborate discussion of the discrete fitting method may be found in \citet{Watvanden13}.

Since Jeans models are computationally much faster to generate than numerical implementations of Schwarzschild models, we are able to evaluate many different model parameters and do a full Bayesian analysis of different models (for example with and without a central massive object). The results presented in this paper were produced using our $C$ implementations of both {\sc nested sampling} \citep{Ski04} and {\sc emcee} \citep{ForHogLan13}. Both codes give similar posterior distributions.

Table \ref{tab:models} gives a summary of the different models fitted to different data sets. We can treat the LOS and PM data as two independent data sets, and as such, we can use the results of one to check the results of the other. We do not attempt to calculate the cross terms between the LOS data and the proper motions, since the number of overlapping stars is so small that doing so would not improve the constraints on the best model. 
    
\subsection{Multi-gaussian expansion of the surface brightness profile}
For the prediction of the second moments of the velocity with Jeans models, accurate photometry is essential. An MGE expansion of the light is convenient, since the deprojection can be done analytically so that the calculation of the gravitational potential requires the evaluation of only one numerical quadrature. Even more so, under certain reasonable (but still ad-hoc) assumptions on the velocity ellipsoid, a solution of the Jeans equations can be calculated with only one numerical quadrature \citep[e.g.][]{Cap08}.

In this paper we use the same MGE expansion as the one used by vdB06, consisting of fourteen components fitted to the smooth surface brightness profile of the cluster derived by \citet{NoyGeb06}.
  
\subsection{Velocity anisotropy and mass density}
To predict second moments of the velocity, a Jeans model requires, besides a light distribution, the velocity anisotropy and a distribution for the mass density. The velocity anisotropy is assumed to be aligned with the cylindrical coordinate system, so that $\overline{v_Rv_z} = 0$ and the velocity anisotropy in the meridional plane $\beta_z = 1 - \frac{\overline{v_z^2}}{\overline{v_R^2}}$ is the only remaining free parameter. 

 The mass density is also described by an MGE, which we obtain by multiplying each component of the luminous MGE by a mass-to-light ratio. Because of computational convenience, an IMBH is approximated by a Gaussian with a very small width of 0.006 arcsec ($\simeq 3\times 10^{-4}$ pc); we have tested that a ten times smaller width does not change our results.

\begin{table*}
\begin{center}
\begin{tabular}{cccccccc}
Model no. & Data & BH mass & $\beta_z$ & Incl. & Free gaussians & Notes\\ 
(1) & (2) & (3) & (4) & (5) & (6)  \\ 
\hline
1 & LOS & $2321\pm1091$  & 0.  & 60. & 4  \\
2 & LOS & $2411\pm1066$ & free & 60. & 5  \\
3 & PM & $1315\pm1015$ & 0. & 60. & 5 \\
4 & PM & $2098\pm1245$ & 0. & 60. & 5& Fitted dynamical center\\
5 & LOS+PM & $2369\pm948$ & 0. & free. & 5 \\
6 & LOS+PM & $2367\pm987$ & free & free. & 5 \\
\end{tabular}
\caption{Summary of fitted models: (1) Model number, as used in the text. (2) Data used: line-of-sight velocities (LOS), proper motions (PM) or both. (3) Best fit black hole mass (in M$_\odot$). (4) Anisotropy parameter $\beta_z$, if not fixed to zero. (5) Inclination in degrees. (6) The number of gaussian components that were left free during the fitting.}\label{tab:models}
\end{center}
\end{table*}

\section{Best-fit parameters}\label{sec:bestfit}
\subsection{Mass-to-light ratio profile}
\label{sec:ml}
\begin{figure*}
  \begin{center}
    \includegraphics[angle=0,width=.48\textwidth]{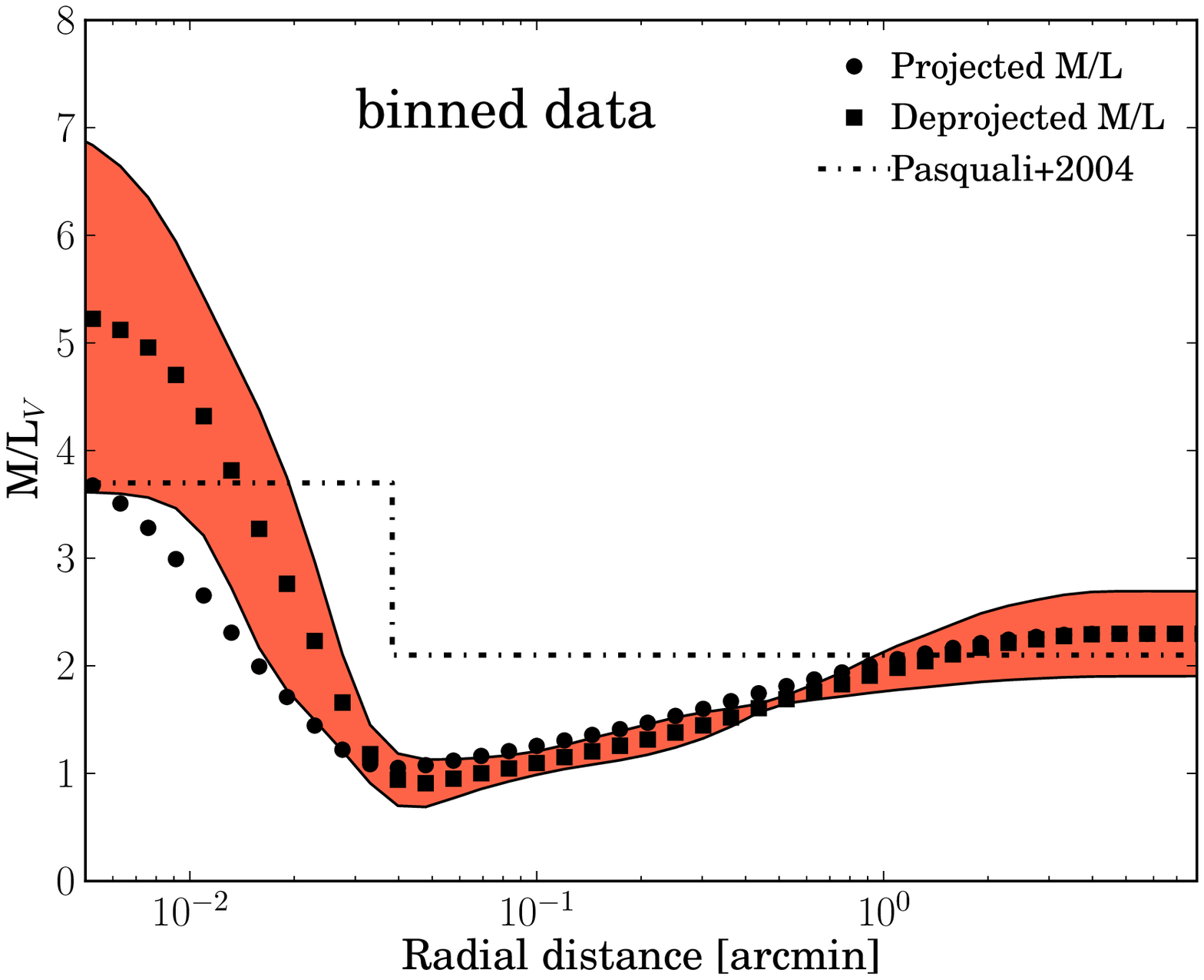}  
    \includegraphics[angle=0,width=.48\textwidth]{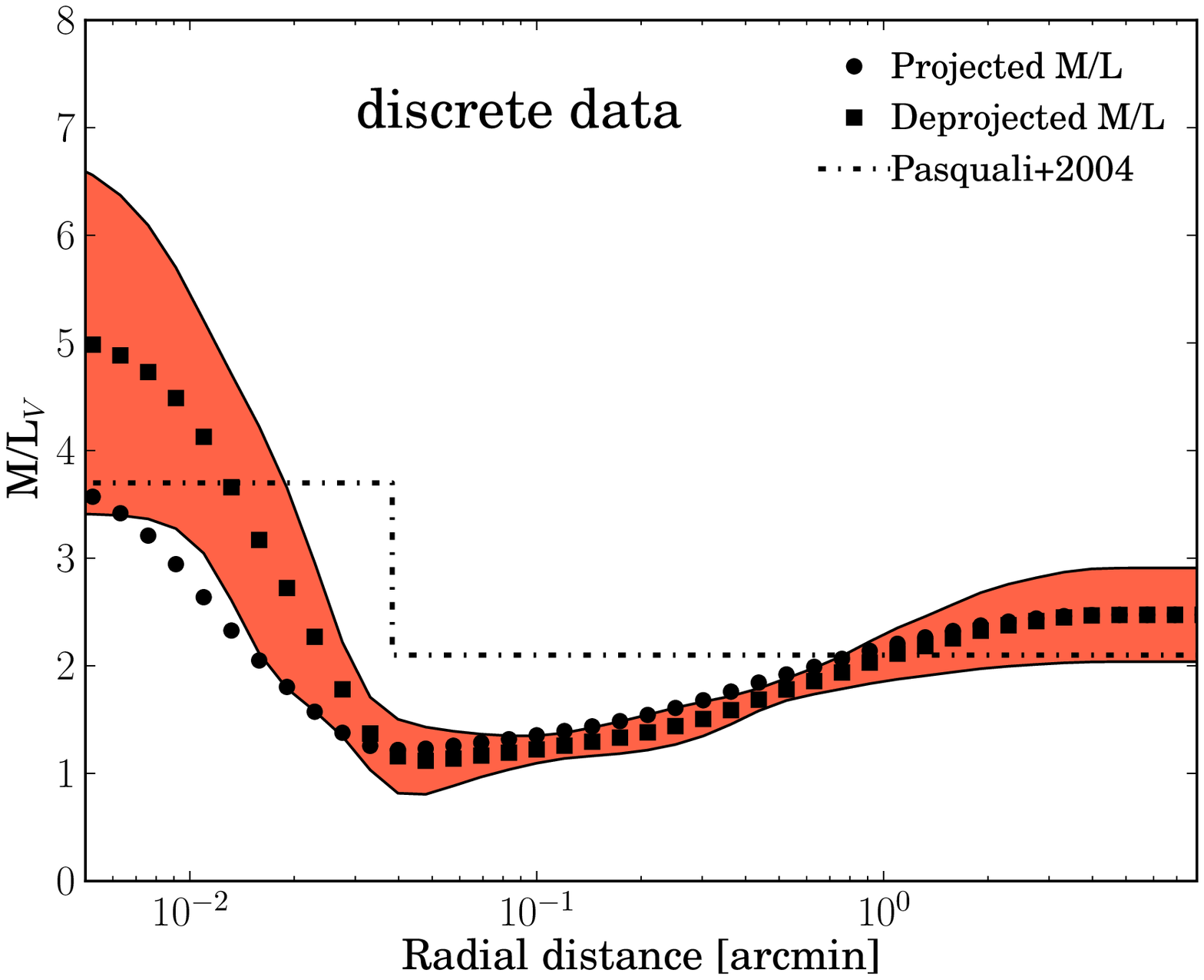}
  \end{center}  
  \caption{The mass-to-light (M/L) profile of M15 (based on model 6), as a function of distance from its center. Left: fitting binned velocity data, right: fitting discrete data. The deprojected and projected profiles are given by squares and circles, respectively. The 1-$\sigma$ confidence interval for the deprojected profile is given by the red area. The measurements from \citet{PasDeMPul04}, based on Michie-King models, are shown with a dashed line.} 
  \label{fig:ml_m15_6}
\end{figure*}
As a first test, we measure the global mass-to-light ratio (M/L). We have left the global velocity anisotropy $\beta_z$ as a free parameter during the fit, but the anisotropy quickly converges to zero (meaning that the velocity ellipsoid is isotropic in the meridional plane). As the light profile is parametrized as a sum of gaussians with different widths, we can give some gaussians a different M/L, instead of forcing a globally constant M/L. As there are 14 gaussians, we leave each of the first central 3 gaussians free (M/L between 0 and 10) together with the 6th and 10th gaussian. The outer gaussians are fixed to the value of the 10th gaussian, and the other gaussians are interpolated linearly in $\log(r)$-space. 

Fig.\ \ref{fig:ml_m15_6} shows the dynamical M/L profile of the cluster in the V-band measured with binning (left panel) and without binning (right panel), based on model 6. The outer profile is almost the same as the one found by vdB06, and is consistent with the previous estimate from \citet{PasDeMPul04}, based on Michie-King models. The increase of M/L toward the center confirms the idea that the center of the cluster is dominated by relatively dark objects: white dwarfs, neutron stars and/or black hole(s). 

\begin{figure*}
  \begin{center}
    \includegraphics[angle=0,width=.98\textwidth]{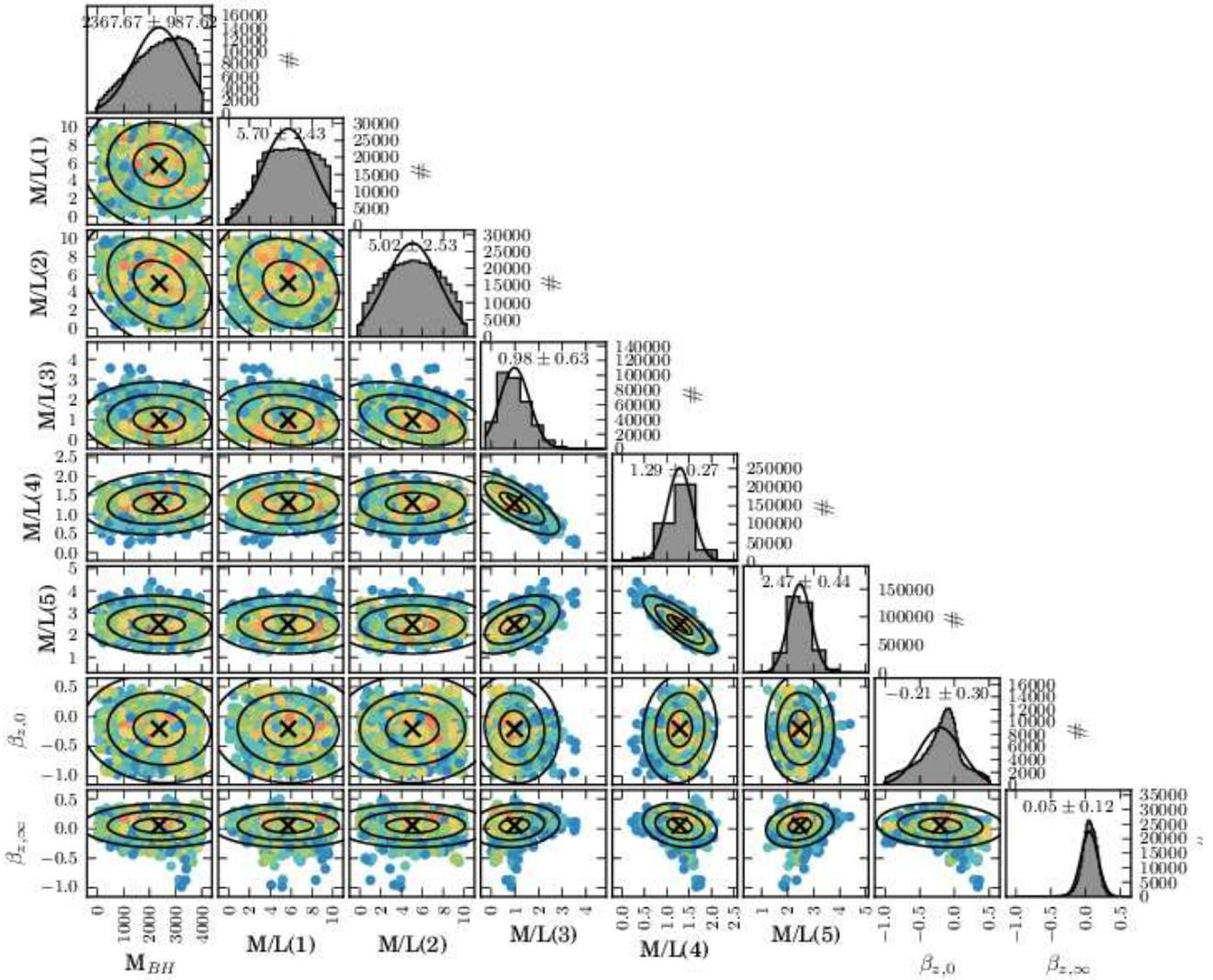} 
  \end{center}  
  \caption{Posterior distribution functions of M/L and $\beta_z$ variables used in model 6. Although all MCMC output points after the burn-in phase were used in the analysis, we only plot the last 2000 points here. The points are coloured by their likelihood, with red colours indicating a high likelihood and blue a low likelihood. The mean and 1, 2 and 3-$\sigma$ error ellipses are overplotted on these points. The histograms show the posterior distribution for a variable after projection over all other variables and the gaussian with equivalent dispersion and mean.} 
  \label{fig:m15_post}
\end{figure*}

In Fig. \ref{fig:m15_post} we show the Markov chain Monte Carlo posterior distributions based on the same model 6 without binning. The points in this figure are coloured by their likelihood, and the projected distribution histograms, with a gaussian with the same mean and width, are shown in this figure as well. This figure shows that we have chosen the priors on the mass-to-light ratios for the determination of the mass density in our dynamical models sufficiently wide. 

\subsection{Velocity anisotropy profile}
\begin{figure}
  \begin{center}
    \includegraphics[angle=0,width=.48\textwidth]{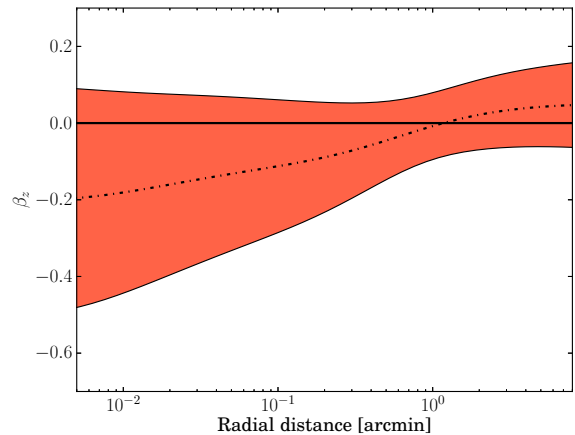}
  \end{center}  
  \caption{Velocity anisotropy $\beta_z$ in the meridional plane. While the inner parts are isotropic, towards the outskirts the stars move on more tangential orbits.} 
  \label{fig:beta_m15_gh_2}
\end{figure}
We found that, when left free, the global velocity anisotropy parameter $\beta_z$ goes almost exactly to zero, meaning that the motions in the meridional plane are isotropic. However, a large contribution to this parameter is coming from stars in the inner parts, and it may be that the inner collapsed part is decoupled from the outer parts. To test this, we fitted the cluster again (with a black hole), and left the anisotropy of the inner Gaussians semi-free, i.e., we parametrize the orbital anisotropy through a modified Osipkov-Merritt profile \citep{Osi79,Mer85}:
\begin{eqnarray}
\beta_z(R) = \beta_0 + \frac{\beta_{\infty} - \beta_0}{1 + \left(\frac{R_\beta}{R} \right)^2}, 
\end{eqnarray}
where $\beta_0,\beta_{\infty}$ and $R_\beta$ are the central velocity anisotropy, asymptotic anisotropy, and turn-over radius. This parametrization thus requires only three free parameters for the whole MGE expansion.
 
In Figure \ref{fig:beta_m15_gh_2} we show the anisotropy of the cluster as a function of radius after fitting LOS and PM data (model 6). The velocity anisotropy is found to be still consistent with isotropy, but inside 1' becomes more negative, i.e. the orbits become more tangential. This is consistent with what is predicted by theoretical models of core collapse, for which one finds isotropy throughout the inner parts of the cluster and radial motions at large radii \citep{Tak95}. 

\subsection{Inclination}
\begin{figure}
  \begin{center}
    \includegraphics[angle=0, width=.48\textwidth]{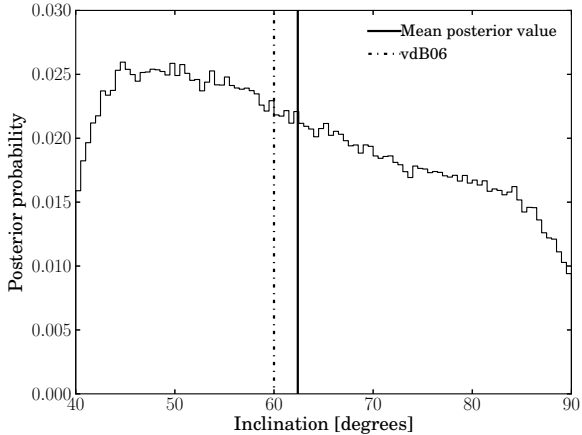}
  \end{center}  
  \caption{The probability density distribution of the inclination of the cluster after projecting all other parameters (M/L, $\beta_z$, black hole mass). The mean value and standard deviation ($62 \pm 14\degree$) show that the inclination is  consistent with the inclination of $60 \pm 15\degree$ found by vdB06, although we note that the posterior distribution is non-gaussian.} 
  \label{fig:inclination}
\end{figure}

Fig\ \ref{fig:inclination} shows the probability distribution function for the inclination, after projection over all other variables (all valid M/L ratios, anisotropies and black hole masses.) The mean resulting inclination ($62 \pm 14\degree$) is consistent with the determination of vdB06 ($60 \pm 15\degree$), although we note that the posterior distribution is highly non-gaussian and not very constraining. The mean inclination coincides with the inclination which we inferred by comparing the mean $y$-PM with the mean LOS (for axisymmetric systems $\overline{v_{los}} = \tan(i) \overline{v_{y}}$, where $i$ is the inclination). Since the determined inclination is close to the one assumed in the other models, the central dark mass is not affected by letting the inclination vary freely.

\section{The central dark mass of M15}
\label{sec:bh}
\begin{figure}
  \begin{center}
    \includegraphics[width=.48\textwidth, angle=0]{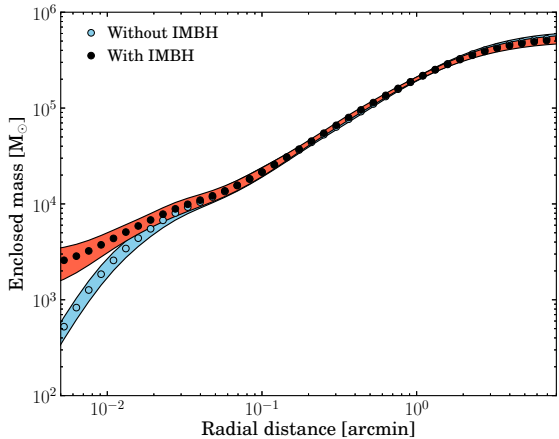}
  \end{center}  
  \caption{The enclosed mass profile of M15 (in solar mass) as a function of radius (in arcmins) for the best model fits without (open circles, blue) and with (closed circles, red) a black hole. Note that the enclosed mass beyond 0.015$^\prime$ is independent of the assumption of an IMBH.} 
  \label{fig:bh_m15_5}
\end{figure}
Our best fitting model contains a dark central gaussian component with a mass of $2\pm 1\times 10^3$ M$_\odot$. However, the difference in Bayesian evidence between model fits with and without this component is less than unity, meaning that there is only a slight preference for a model with a dark central component. In order to prove that this dark component represents a black hole in M15, we should show that the mass in the center of M15 is so concentrated that there is no other solution possible. In practice, this is not possible, since it requires the measurement of velocities at a few Schwarzschild radii from the black hole. However, a concentrated dark central mass would dominate the kinematics within the sphere of influence. The sphere of influence of the black hole would be
\begin{eqnarray}
r_{bh} = \frac{G M}{\sigma^2} \simeq 0.07\;\mbox{pc} \simeq 0.023' 
\end{eqnarray}
In Fig.\ \ref{fig:bh_m15_5} we see that we only barely resolve the sphere of influence, but that both fits give a more or less equal mass within 0.015' ($\simeq  0.046$ pc).  Apparently, at this radial range, we can measure the slope of the potential accurate enough to constrain a high central density, though we can not resolve far enough inside to distinguish between a point-like mass and a smooth mass distribution -- this would require more kinematic measurements of stars close to the centre of the cluster.

\section{Discussion and Conclusions}\label{sec:conc}
We have fitted dynamical models to the kinematic data of M15. Models which were fitted independently to LOS and PM data result in a very high density at the center of the cluster. Because of mass-segregation, the centre of the cluster likely contains a large conglomeration of stellar remnants, small black holes, neutron stars and white dwarfs. We have however seen that the {\it M/L} profile is relatively constant throughout the core - and only in the very center is extra mass required. Does the fact that the {\it M/L} profile has a central plateau, but may require the addition of a compact gaussian, suggest that there is an IMBH in M15?
 
So far, all evidence goes against a black hole: radio observations of M15 have not shown evidence for the presence of a black hole -- there is a large dependence on the assumptions for the presence of interstellar matter (ISM) and the accretion rates, but upper limits on the mass of the IMBH have been set as low as 440\Msun \citep{Mac04} and 1000\Msun \citep{BasGebGos08}. But even with conservative assumptions for the accretion rate and ISM, it is possible that there is an IMBH.
 
From theoretical considerations, an IMBH is not expected. \citet{BauHutMak03} predict the slope of the mass density of stellar remnants. 
\begin{figure}
  \begin{center}
    \includegraphics[angle=0, width=.48\textwidth]{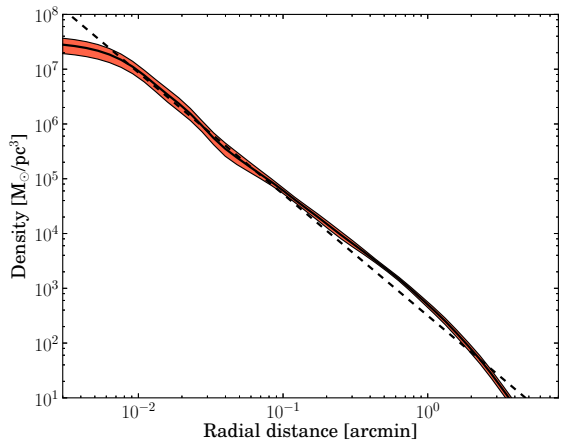}
  \end{center}  
  \caption{The inferred mass-density profile of M15 as a function of radius shown with a solid line. The surrounding (red) region shows the 1-$\sigma$ error on this profile. The dashed line shows the expected power-law mass-density profile for dark remnants (white dwarfs and neutron stars) from \citet{BauHutMak03} (scaled in amplitude to match the inferred profile around 0.1').} 
  \label{fig:density}
\end{figure}
In Fig\ \ref{fig:density} we show our inferred mass-density profile. The dashed line follows the slope of the mass density of stellar remnants, found in N-body simulations \citep{BauHutMak03}, though scaled in amplitude to match the inferred profile around 0'.1, so that: 
\begin{eqnarray}
\rho(r) = 8754 \left(\frac{r}{arcmin}\right)^{-2.22} \frac{M_\odot}{\mbox{arcmin}^3.}
\end{eqnarray}
The simulated profile matches our inferred density profile, constrained by LOS and PM data, very well. However, in the very center (and in the outskirts where main sequence stars are the main contributors to the mass) the two profiles diverge. Since the black hole is assumed to be a point source, it does not show up in the density plot. It may be that the inferred IMBH is just a consequence of too coarse sampling in the center. To test this, we use the theoretical mass-density profile of Baumgardt et al. (2003) to estimate how much mass in our best-fit model may be attributed to a black hole.  Under the assumption that this density profile continues to the centre of the cluster, the mass inside 0.01' is $\sim 3.9 \times 10^3 M_\odot$. If we use the mass in the dark central gaussian component and our derived M/L profile, we find an enclosed mass within 0.01' of $\sim 4.4 \pm 1.4 \times 10^3 M_\odot$. 

Although the current resolution in the center is too low to make a definitive statement, it suggests that the inferred black hole mass might be a consequence of poor sampling of the density profile in the central parts. {\it This means that, even though we find a higher mass density in the center than compared to previous dynamical modelling, the central mass density is in agreement with theoretical predictions and provides no evidence for an IMBH in M15.}
The match between our derived M/L profile without binning and the previously-determined M/L profiles suggests that dynamical modelling of discrete stellar systems does not require any binning. This not only leads to more accurate fits but also suppresses biases in the central mass determinations due to loss of resolution when data are binned. The close agreement between the power law slope of the density as predicted by numerical simulations and the power-law slope measured from our dynamical modelling suggests that we really do understand the dynamics in the inner parts of the core-collapsed globular cluster M15.

There are several problems to which the discrete kinematics fitting method can be applied: simultaneous modelling of member and interloper stars for Milky Way globular clusters and dwarf spheroidals \citep[see also][]{Watvanden13}, modelling different kinematic tracer populations without applying hard cuts to the data \citep[see also][]{WalPen11}, or simultaneous modelling of both individual stars and aggregates of stars in IFU observations.

\section*{Acknowledgments} 
We thank the referee for constructive comments. This work was supported by Sonderforschungsbereich SFB 881 "The Milky Way System" (subproject A7) of the German Research Foundation (DFG). 

\bibliographystyle{mn2e}
\bibliography{paper}

\end{document}